# Fully Automatic Brain Tumor Segmentation using a Normalized Gaussian Bayesian Classifier and 3D Fluid Vector Flow

*Tao Wang, Irene Cheng and Anup Basu*
Department of Computing Science, University of Alberta

**ABSTRACT**

Brain tumor segmentation from Magnetic Resonance Images (MRIs) is an important task to measure tumor responses to treatments. However, automatic segmentation is very challenging. This paper presents an automatic brain tumor segmentation method based on a Normalized Gaussian Bayesian classification and a new 3D Fluid Vector Flow (FVF) algorithm. In our method, a Normalized Gaussian Mixture Model (NGMM) is proposed and used to model the healthy brain tissues. Gaussian Bayesian Classifier is exploited to acquire a Gaussian Bayesian Brain Map (GBBM) from the test brain MR images. GBBM is further processed to initialize the 3D FVF algorithm, which segments the brain tumor. This algorithm has two major contributions. First, we present a NGMM to model healthy brains. Second, we extend our 2D FVF algorithm to 3D space and use it for brain tumor segmentation. The proposed method is validated on a publicly available dataset.

*Index Terms—* brain tumor segmentation, vector flow

## 1. INTRODUCTION

Brain tumor segmentation from MRIs is an important task for neurosurgeons, oncologists and radiologists to measure tumor responses to treatments [1]. Manual segmentation takes considerable time and is prone to error. Therefore, automatic brain tumor segmentation methods have been highly desirable in recent decades. However, automatic brain tumor segmentation is a very challenging task due to many factors. For instance, different types of brain tumors have large variations in sizes, shapes, locations and intensities.

Many brain tumor segmentation methods have been proposed. They can be classified into two categories: training based methods and non-training based methods.

Training based methods often use brain tumor images or healthy brain images to train a segmentation model and use other brain tumor images to test the model. Warfield *et al.* [2] presented an adaptive, template moderated (ATM), spatially varying statistical classification (SVC) method for brain tumor segmentation. Kaus *et al.* [3] extended this idea and proposed a classification algorithm to segment brain MRIs into five different tissue classes (background, skin, brain, ventricles, and tumor). The algorithm was validated in a dataset of 20 patients with low-grade gliomas and meningiomas. Corso *et al.* [4] integrated a Bayesian formulation into the SWA (segmentation by weighted aggregation) algorithm to segment GBM. The SWA algorithm used voxel intensities in a neighborhood to compute an affinity between the respective voxels. However, it is not clear whether this method can segment other types of brain tumors.

Non-training based methods, on the other hand, do not have any training process. Since the training or learning process can provide valuable information for brain tumor segmentation, only a handful of non-training based methods have been proposed in the literature. Phillips *et al.* [5] used a fuzzy C-means (FCM) clustering algorithm to segment brain tumors from normal brain tissues. FCM is similar to the k-means algorithm [14] for unsupervised clustering but allows labels to be "fuzzy." One of the drawbacks of this method is that human interaction is required to segment brain tumor. Karayiannis *et al.* [6] proposed a fuzzy algorithm for 2D brain tumor segmentation. Ho *et al.* [7] incorporated region competition into an active contour model for brain tumor segmentation.

In this paper, a brain tumor segmentation method is presented and validated on a publicly available brain tumor segmentation repository [2-3, 8] with ten patients. This method: 1) is fully automatic; 2) works in 3D; and 3) requires only T1 MRIs. In our method, brain MR images are pre-processed with the software MIPAV [9, 10]. After this pre-processing procedure, there are three stages. In the first stage, a "Normalized" Gaussian Mixture Model (NGMM) is proposed and estimated by Expectation-Maximization (EM) based on the ICBM452 brain atlas [11]. NGMM is then used to model healthy and normal brain tissues. In the second stage, the ICBM Tissue Probabilistic Atlases [12] are utilized to obtain the prior probabilities of different brain tissues. After that, a Gaussian Bayesian Classifier based on the NGMM and the prior probabilities of different brain tissues is exploited to acquire a Gaussian Bayesian Brain Map (GBBM) from the test 3D brain MR images. GBBM is further processed to highlight the brain tumor and initialize a Fluid Vector Flow (FVF) algorithm. In the last stage, FVF is used to segment the brain tumor. There are two major contributions in this paper. First, we introduce a new NGMM algorithm to model healthy and normal brain. This model can be easily modified for modeling other tasks in various application domains. Second, we extend our 2D FVF algorithm [13] to 3D space and use it for automatic brain tumor segmentation. One drawback of our previous 2D FVF algorithm was that an initial contour was needed to start the vector flow evolution. In this paper, we take advantage of the GBBM to provide an initial position of a brain tumor to the 3D FVF algorithm to make this process fully automatic.

The rest of this paper is organized as follows. Section 2 introduces the proposed brain tumor segmentation method. Experimental results are reported in Section 3, before the work is concluded in Section 4.

## 2. PROPOSED METHOD

MR images must be pre-processed before further processing and analysis. The details of the pre-processing steps are described in Section 3.3.

### 2.1. Normalized Gaussian Mixture Model and Gaussian Bayesian Brain Map

In this section a new Normalized Gaussian Mixture Model (NGMM) is proposed. The basic idea of GMM is to use multiple Gaussian distributions to model multiple brain tissues such as gray matter (GM), white matter (WM), and cerebrospinal fluid (CSF). To utilize GMM, the Gaussian distributions of the brain atlas and the brain tumor dataset must be aligned correctly. Unfortunately, this is not true in our dataset. For example, Fig. 1 shows the histogram of the ICBM452 atlas and the histogram of the MRIs of Patient #1. The intensity regions of CSF, GM, and WM are marked. The intensity range of the atlas is [0, 712] while the intensity range of the MRIs of Patient #1 is [0, 567]. Note that contrast stretching technique cannot align the Gaussian distributions in case a "long tail" exists (see Fig. 1).

To align the Gaussian distributions without eliminating potential brain abnormalities, we define the normalized intensity value $P_{out}$ as:
$$P_{out} = P_{in} / m \quad (1)$$
where $m$ is the mean image intensity.

With this definition, the mean image intensity is normalized to 1 and the majority of image intensity is stretched to a range around the mean value. The Gaussian distributions are also aligned so that GMM can be utilized.

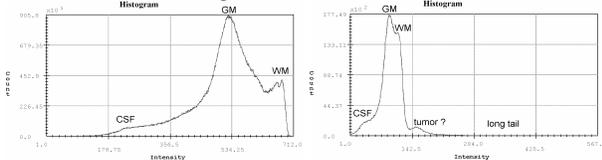

**Fig. 1**. (Left) histogram of ICBM452 brain atlas. (Right) histogram of the MR images of patient #1.

We present a Normalized Gaussian Mixture Model (NGMM) to model the healthy or normal brain in the ICBM452 atlas. NGMM estimates a Gaussian Mixture Model based on the normalized image intensities as:
$$p(x_i) = \sum_{k=1}^{K} p(k) p(x_i | k) \quad (2)$$
where $x_i$ is a voxel in the image, $p(k)$ is the prior probability, and $p(x_i | k)$ is the conditional probability density, which is define as:
$$p(x_i | k) = N(x_i; \mu_k, \sigma_k) = \frac{1}{\sigma_k \sqrt{2\pi}} \exp(-\frac{(x_i - \mu_k)^2}{2\sigma_k^2}) \quad (3)$$

where $\mu_k$ is the mean and $\sigma_k$ is the standard deviation of the Gaussian $N(x_i; \mu_k, \sigma_k)$.

**Prior probabilities**

The ICBM Tissue Probabilistic Atlases are utilized to obtain the prior probabilities of different brain tissues. At a given voxel $x_i$, the prior probability $p(k, x_i)$ is defined as:
$$p(k, x_i) = \xi(k, x_i) / \sum_{k=1}^{K} \xi(k, x_i) \quad (4)$$

where $\xi(k, x_i)$ is image intensity of different brain tissues in the three ICBM Probabilistic Atlases (CSF, GM, or WM).

**Gaussian Bayesian Brain Map**

At a given 3D location $(u, v, w)$ the conditional probability $p(k | \xi)$ can be calculated by Bayes' Theorem:
$$p(k | \xi) = p(k) \cdot p(\xi | k) / \sum_{k=1}^{K} p(k) \cdot p(\xi | k) \quad (5)$$

The correlation coefficient $CC \in [-1,1]$ at this voxel is then calculated:
$$CC = Cov(\alpha, \beta) / \sqrt{Cov(\alpha, \alpha) \cdot Cov(\beta, \beta)}$$
where $\alpha = (p(CSF | \xi), p(GM | \xi), p(WM | \xi))$, $\beta = (p(CSF), p(GM), p(WM))$, and $Cov$ is covariance.

The correlation coefficient $CC$ can reveal the likelihood of finding a candidate tumor voxel at a given 3D location. When $CC$ is close to -1, it means the intensity of this voxel disagrees with the NGMM and this voxel is probably abnormal or likely to be a tumor voxel. When $CC$ is close to 1, it means that the intensity of this voxel agrees with the NGMM and this voxel is likely to be normal. The correlation coefficient is then used to define the Gaussian Bayesian Brain Map (*GBBM*):
$$CM_{uvw} = \begin{cases} 1.0 - CC_{uvw} & \text{when } CC_{uvw} > 0.0 \\ 0.0 - CC_{uvw} & \text{otherwise} \end{cases} \quad (6)$$

$$GBBM = [a_{uvw}]_{m \times n \times o} \quad \text{where } a_{uvw} = \Omega \cdot CM_{uvw} \quad (7)$$

$CC$ is first mapped to $CM$ in the region of [0, 1]. Then $GBMM$ is calculated based on CM and $\Omega$. $\Omega$ is a scaling parameter that represents the intensity range of *GBBM*. The resulting *GBBM* is a 3D matrix or image with dimension of $m \times n \times o$, $m, n$ and $o \in N$. GBBM reveals the likelihood of finding a candidate tumor in the entire image domain.

### 2.2. Candidate Tumor Region

GBMM is further processed to highlight a candidate tumor region. This region will be used to automatically initialize the 3D Fluid Vector Flow algorithm, which will finally segment the brain tumor. The post-processing stage has six steps: removing boundary voxels, thresholding, morphological erosion, locating the largest 3D region, morphological dilation, and reverse transformation.

Brain tumor MR images are registered to a brain atlas in the pre-processing stage. However, the registered image and the atlas are still different at a few voxels on the brain boundary. Therefore, boundary voxels must be removed from *GBMM*. A threshold $\Psi$ is then applied to *GBMM* to create a Binary Gaussian Bayesian Brain Map (*BGBBM*):

$$b_{uvw} = \begin{cases} 1 & \text{when } a_{uvw} > \Psi \\ 0 & \text{otherwise} \end{cases} \quad (8)$$

$$BGBBM = [b_{uvw}]_{m \times n \times o} \quad (9)$$

We use morphological filters to merge relatively large separated regions. Then the largest 3D region is automatically located. This region represents the candidate tumor. Dilation is then used to restore the region to approximately its original size. Fig. 2 shows the GBMM and the candidate tumor region after the dilation for Patient #1.

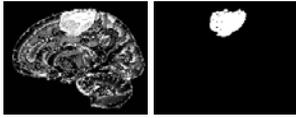

**Fig. 2**. (Left) Gaussian Bayesian Brain Map of the brain. (Right) The candidate tumor region after dilation.

Finally, we apply $T^{-1}$ to the dilated image to transform it back to the original image space.

### 2.3. 3D Fluid Vector Flow

Fluid Vector Flow (FVF) [13] is a snake model. It has been applied to semi-automatic brain tumor segmentation in 2D space. One drawback of the previous 2D FVF algorithm was that an initial contour was needed to start the vector flow evolution. In this paper, the FVF algorithm is extended to 3D space and the candidate tumor region is used to initialize the 3D FVF algorithm to make this process fully automatic.

There are two types of snakes in the literature: parametric [15] and level set snakes [16]. While parametric snakes mainly focus on the 2D domain, level set snakes have been used for 3D segmentation and reconstruction [16]. A level set [17-18] snake is an implicit model, which is not explicitly expressed as a parametric model but is implicitly specified as a level set of a scalar function $\phi$. A 3D surface may be written as:

$$f(x, y, z) = (x(s_1, s_2), y(s_1, s_2), z(s_1, s_2)) \quad (10)$$

$F = F(K)$ is the speed of the surface evolution

$$F^2(f_x^2 + f_y^2 + f_z^2) = 1 \quad (11)$$

where K is the mean curvature. The surface motion is formulated as a Partial Differential Equation (PDE):

$$\frac{\partial \phi}{\partial t} - F(K) |\nabla \phi| = 0 \quad (12)$$

where $\phi(x, y, z, t)$ is a scalar function such that at time $t$ the zero level set of $\phi$ is the surface.

3D Fluid Vector Flow (FVF) is used to segment brain tumors in this paper. It is an extension of our 2D FVF algorithm [13], which had been applied to semi-automatic brain tumor segmentation in 2D space. 3D FVF takes the candidate tumor region obtained in the previous section and automatically segments the brain tumor by taking advantage of the level set method. Chan and Vese [19] pointed out that the evolution of level sets should not always rely on surface normal. The basic idea of FVF is to add a directional component to the external force and keep the normal component. The novelty of FVF lies in the computation of the directional force. When the active surface is within the candidate tumor region, the directional force can push the surface towards the boundary of the tumor. When the surface is close to the boundary of the tumor, the normal force fits the surface to the tumor.

The basic idea of FVF is to add a directional component to the external force and keep the normal component. At a given point $B = (x_1, y_1, z_1)$ on the level set surface, there are two straight lines $L_1$ and $L_2$:

$$L_1 : \frac{x - x_1}{l_1} = \frac{y - y_1}{m_1} = \frac{z - z_1}{n_1} \quad L_2 : \frac{x - x_1}{l_2} = \frac{y - y_1}{m_2} = \frac{z - z_1}{n_2}$$

Where $L_1$ is the surface normal at point $(x_1, y_1, z_1)$ and $L_2$ is the straight line determined by vector $\overrightarrow{AB}$ that start from center $A = (x_0, y_0, z_0)$ and points to $B = (x_1, y_1, z_1)$.

Next, we extend the external energy function in [13] to 3D:

$$E_e(x, y, z) = \chi(f_x + \delta\gamma_x, f_y + \delta\gamma_y, f_z + \delta\gamma_z)$$

where $\chi$ is a normalization operator, $\delta = \pm 1$ (controlling the inward or outward direction, when the surface is "outside" or "inside" the candidate tumor region), and $\gamma$ is the angle between $L_1$ and $L_2$, determined by:

$$\cos\gamma = \frac{l_1 l_2 + m_1 m_2 + n_1 n_2}{\sqrt{l_1^2 + m_1^2 + n_1^2}\sqrt{l_2^2 + m_2^2 + n_2^2}}$$

Details on extending 2D FVF to 3D is not shown here due to space limitations.

### 3. EXPERIMENTS

We tested the proposed method with the SPL Dataset [8]. We used two ICBM atlases, the ICBM452 T1 Atlas [11] and the ICBM Tissue Probabilistic Atlases [12]. The ICBM452 atlas is used to estimate a Normalized Gaussian Mixture Model. The ICBM Tissue Probabilistic Atlases [12] are used to obtain the prior probabilities of GM, WM, and CSF.

We chose software package MIPAV [9-10] for pre-processing. Our pre-processing stage has two steps: skull stripping and registration. Fig. 3 shows the original MR image, the extracted brain, the ICBM452 atlas, and the registered brain of Patient #1 in the dataset.

The Tanimoto Metric [20] is used for quantitative analysis. It is defined as: $TM = |R_X \cap R_G| / |R_X \cup R_G|$, where $R_X$ is the region enclosed by the surface generated by the proposed method, $R_G$ is the region of the ground-truth segmentation provided in the SPL Brain Tumors

Image Dataset, and | | is set cardinality. Table I shows the test results. In some test cases (e.g., case #1 and #2), the accuracy is satisfactory. However, the accuracy is not satisfactory in other test cases (e.g., case #5 and #9). Nevertheless, we notice that the accuracy (0.22-0.88) of our method matches a recent work by Corso *et al.* [4], where the accuracy was in the range of 0.27-0.88.

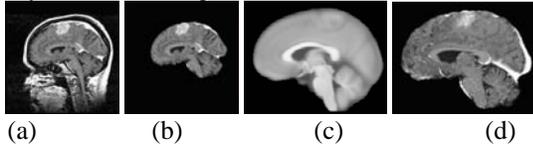

(a)      (b)      (c)      (d)

**Fig. 3**. (a) Original image (b) extracted brain (c) ICBM452 atlas (d) registered brain

**Table I**: Tanimoto Metric (TM) of the proposed method

| Case | Tumor type | Time | TM |
|---|---|---|---|
| 1 | meningioma | 412 sec | 0.88 |
| 2 | meningioma | 423 sec | 0.83 |
| 3 | meningioma | 427 sec | 0.57 |
| 4 | low grade glioma | 424 sec | 0.66 |
| 5 | astrocytoma | 426 sec | 0.22 |
| 6 | low grade glioma | 441 sec | 0.53 |
| 7 | astrocytoma | 439 sec | 0.67 |
| 8 | astrocytoma | 437 sec | 0.57 |
| 9 | astrocytoma | 417 sec | 0.30 |
| 10 | low grade glioma | 449 sec | 0.70 |

Fig. 4 shows the results for Patient #1. The results for other patients are not shown because of space limitations.

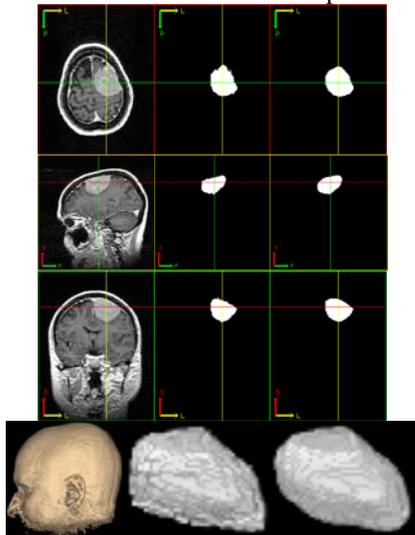

**Fig. 4**. Results of Patient #1. 1st column: brain MR image, 2nd column: ground truth, 3rd column: brain tumor extracted by the proposed method. 1st row: axial view, 2nd row: sagittal view, 3rd row: coronal view, 4th row: volume rendering.

## 4. CONCLUSION AND FUTURE WORK

A new brain tumor segmentation method was presented and validated in this paper. The method is able to segment brain tumors fully automatically by taking advantages of the proposed Normalized Gaussian Mixture Model (NGMM) and 3D Fluid Vector Flow (FVF) algorithms. Utilizing multiple MRI protocols often provide easier and better segmentation [5]. In the future, we will further improve this method by considering multiple modalities (e.g., T2).

## 5. ACKNOWLEDGEMENT

The authors thank Drs. Simon Warfield, Michael Kaus, Ron Kikinis, Peter Black and Ferenc Jolesz for sharing the brain tumor database [2-3].